\documentclass[
superscriptaddress,
preprint,
amsmath,amssymb,
aps,
prb,
]{revtex4-2}

\usepackage[dvipdfmx]{graphicx}
\usepackage{dcolumn}
\usepackage{bm}
\usepackage{amsmath}
\usepackage{soul}
\usepackage{color}
\usepackage[
  dvipdfmx   
  ]{hyperref}

\usepackage{ulem}

\newcommand {\tn}{T_{\rm{N}}}

\definecolor{green}{rgb}{0,0.6,0.1}

\begin{document}


\title{Anomalous metallic states at magnetic interfaces in an antiferromagnetic topological insulator candidate DyPtBi with ferroquadrupolar order}

\author{Kentaro Ueda}
 \affiliation{Department of Applied Physics, University of Tokyo, Tokyo 113-8656, Japan}

\author{Hiraku Saito}
 \affiliation{Institute of Solid State Physics, University of Tokyo, Kashiwa 277-8561, Japan}

\author{Zen Ogame}
 \affiliation{Department of Quantum Matter, AdSE, Hiroshima University, Higashi-Hiroshima 739-8530, Japan}
 
\author{Isao Ishii}
 \affiliation{Department of Quantum Matter, AdSE, Hiroshima University, Higashi-Hiroshima 739-8530, Japan}

\author{Ruo Hibino}
 \affiliation{Department of Physics, Kobe University, Kobe, Hyogo 657-8501, Japan}

\author{Tatsuya~Yanagisawa}
 \affiliation{Department of Physics, Hokkaido University, Sapporo 060-0810, Japan}


\author{Shinichi Itoh}
 \affiliation{Institute of Materials Structure Science, High Energy Accelerator Research Organization, Tsukuba 305-0801, Japan}
 \affiliation{Materials and Life Science Division, J-PARC Center, Tokai 319-1106, Japan}

\author{Taro Nakajima}
 \affiliation{Institute of Solid State Physics, University of Tokyo, Kashiwa 277-8561, Japan}
 \affiliation{Institute of Materials Structure Science, High Energy Accelerator Research Organization, Tsukuba 305-0801, Japan}
 \affiliation{RIKEN Center for Emergent Matter Science (CEMS), Wako 351-0198, Japan}
 
\author{Yoshinori Tokura}
 \affiliation{Department of Applied Physics, University of Tokyo, Tokyo 113-8656, Japan}
 \affiliation{RIKEN Center for Emergent Matter Science (CEMS), Wako 351-0198, Japan}
 \affiliation{Tokyo College, University of Tokyo, Tokyo 113-8656, Japan}





\date{\today}

\begin{abstract}
Antiferromagnetic topological insulators provide a fertile platform where symmetry-breaking magnetic order is intertwined with topological electronic states.
In particular, magnetic domain walls have attracted much attention, as they can be easily controlled by external fields as in general magnets, and moreover, host nontrivial electronic states distinct from those in bulks and sample surfaces.
Here, we report a new antiferromagnetic topological insulator candidate DyPtBi, which hosts conductive magnetic domain walls controllable by uniaxial stress and magnetic field.
We find that the resistivity exhibits abrupt increase upon the magnetic and structural phase transition.
Concomitantly, the transverse ultrasonic mode shows remarkable softening of 6 \%, indicating that the Dy $4f$ ferroquadrupolar order plays a vital role in the phase transition.
Furthermore, we reveal by neutron experiments that applying compressive uniaxial stress aligns the magnetic domain state, leading to the strong resistivity enhancement of 14 \% while eliminating the conductive magnetic domain walls.
These findings demonstrate that DyPtBi exhibits topological electronic states entangled with multipolar degrees of freedom, providing a promising route for \textit{in situ} control of topological properties.
\end{abstract}

\maketitle


\section{Introduction}

The band topology is one of the central topics in contemporary condensed-matter physics.
Increasing attention has been directed toward composite topological materials (CTMs), where topological electronic states are coupled with other degrees of freedom.
For instance, magnetic topological insulators (TIs) show quantum anomalous Hall effect~\cite{2013ScienceChang} and axion insulating state~\cite{2017NMMogi}, offering the possibility of tuning topological properties via manipulating magnetic order by external fields.

\begin{figure*}
\includegraphics[width=1.0\columnwidth]{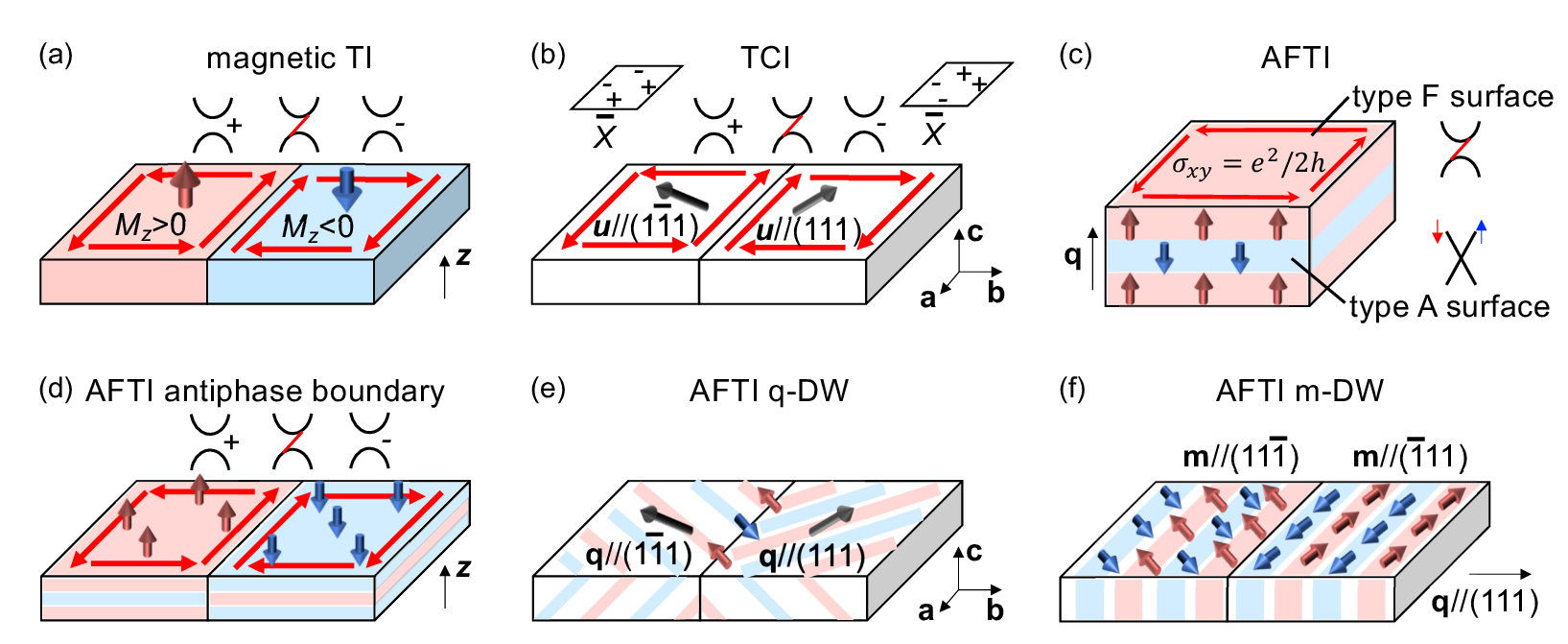}
\caption{\label{fig0}
Schematic pictures of electronic states at domain walls (DWs) for (a) magnetic topological insulator (TI) and (b) topological crystalline insulator (TCI). The red and blue thick arrows in (a) indicate the magnetization direction in respective domains. The thick black arrows in (b) denote the direction of the atomic displacement $\boldsymbol{u}$ in respective domains. The top panels of these figures schematically show the surface band structures and surface Brillouin zone (001) for respective domains and their boundaries. The $+$ and $-$ indicate the signs of the Chern number. The red thin arrows indicate the chiral edge currents.
(c) Schematic picture of antiferromagnetic topological insulator (AFTI). $\boldsymbol{q}$ denotes the magnetic propagation vector. The red (blue) arrows denote the magnetic moments $\boldsymbol{m}$ pointing upward (downward). Two types of surfaces are shown; type F (ferromagnetic) and type A (antiferromagnetic) surface, each of which has the different electronic state due to the breaking or preservation of $S$ symmetry (see text).
(d) Schematic picture of antiphase boundary in AFTI. The red arrows located at domain boundaries indicate the chiral edge current.
Schematic pictures of DWs between (e) domains with different $\boldsymbol{q}$ (q-DW) and (f) with different $\boldsymbol{m}$ (m-DW), respectively. The thick black arrows in (e) indicate $\boldsymbol{q}$ in respective domains.
}
\end{figure*}

Among them, one-dimensional chiral edge channels at magnetic domain walls (DWs) in magnetic TIs have received considerable interest~\cite{2017ScienceYasuda}. Since the sign of the Chern number changes across the domain boundary in accordance with the magnetization direction, gapless states are realized at DWs [Fig.~\ref{fig0}(a)].
Similar nontrivial DW states have also been discussed in the context of topological crystalline insulators (TCIs)~\cite{2012NCHsieh}. For example, SnTe, a prototypical TCI, undergoes a structural distortion along [111] or its equivalent directions ($\boldsymbol{u}//\langle 111\rangle $) at low temperatures~\cite{1975JPSJIizumi}. The rhombohedral distortion breaks the mirror symmetry, leading to the finite mass of the Dirac surface states. The sign of the mass $m_{i}$, corresponding to the sign of the Chern number, is determined by $m_{i}\propto (\boldsymbol{u}\times \boldsymbol{K}_{i})\cdot \hat{z}$ where $\boldsymbol{K}_{i}$ is the momentum of Dirac points and $\hat{z}$ is the surface normal.
The top of Fig.~\ref{fig0}(b) shows the sign of the Dirac mass, which are located near $\bar{X}$ in the surface Brillouin zone (001), for $\boldsymbol{u}//(111)$ and $(1\bar{1}1)$ domains. Notably, when the sign of $m_{i}$ is opposite between adjacent domains, chiral edge states can emerge at their boundaries [Fig.~\ref{fig0}(b)], analogous to those in magnetic TIs.
Therefore, CTMs host DWs that are free from surface-contamination and amenable to external-field control, as in conventional magnetic or dielectric materials, while supporting topological edge states.

An antiferromagnetic topological insulator (AFTI) is also a representative CTM which is characterized by a nontrivial $Z_{2}$ invariant defined under the combined symmetry operation $S=\Theta T_{\rm 1/2}$ consisting of time-reversal symmetry $\Theta $ and a primitive-lattice translational symmetry $T_{\rm 1/2}$~\cite{2010PRBMong}.
Of particular interest is the surface states. As shown in Fig.~\ref{fig0}(c), an antiferromagnetic surface, referred to as the type A surface in the literature~\cite{2010PRBMong}, preserves the $S$ symmetry and hence hosts the surface states analogous to those of TIs. On the other hand, the two-dimensional surface states are gapped on a ferromagnetic surface (type F) where the $S$ symmetry is broken.
In other words, AFTI can be regarded as a stack of half-integer quantum Hall layers with alternating magnetization directions. Consequently, chiral edge currents are expected at step edges and antiphase boundaries [Fig.~\ref{fig0}(d)] which are directly observed in MnBi$_2$Te$_4$~\cite{2020PRLSass}.
A much richer variety of magnetic domains are present in antiferromagnets.
Let us consider a prototypical antiferromagnet such as cubic NiO, characterized by a magnetic propagation vector $\boldsymbol{q}//\langle 111\rangle $~\cite{1960JAPSlack}. It hosts DWs between different $q$ domains, termed here as q-DWs [Fig.~\ref{fig0}(e)], on which the spin configuration is categorized to type A.
Since the $S$ operation for each domain contains a shift of the different direction, each electronic structure cannot be adiabatically connected and hence nontrivial states may be realized at DWs, similar to TCI [Fig.~\ref{fig0}(b)].
Furthermore, as illustrated in Fig.~\ref{fig0}(f), there are domain boundaries associated with a change in the easy axis of magnetic moments $\boldsymbol{m}$, which are referred to as m-DWs. Although the $S$ symmetry is broken in this case, a number of exotic states, including spin-polarized flat bands~\cite{2021PRBPetrov} and topologically protected zero-line modes~\cite{2023PRBLiang}, are theoretically proposed at DWs.
Not only atomically-thin DWs mentioned above, but also more complex DWs such as Bloch- or Neel-type can be realized.
Thus, magnetic interfaces exhibit a diverse range of spin configurations, potentially giving rise to novel topological phenomena absent in both surfaces and bulks.

To date, only a scarce number of AFTI candidates have been identified, such as MnBi$_2$Te$_4$~\cite{2019NatureOtrokov} and EuIn$_2$As$_2$~\cite{2021NCRiberolles}.
They exhibit a number of remarkable phenomena including various topological phases depending on odd/even-number layers~\cite{2019ScienceDeng}, field-controlled Dirac surface states~\cite{2021NCRiberolles}, and quantum metric nonlinear Hall effect~\cite{2023ScienceGao}.
However, as for the magnetic domain states, MnBi$_2$Te$_4$ has only two types of domains, resulting in only antiphase boundaries as shown in Fig.~\ref{fig0}(d). 
Thus, expanding the range of candidate materials and functionalities is highly desired.

\begin{figure}
\includegraphics[width=1.0\columnwidth]{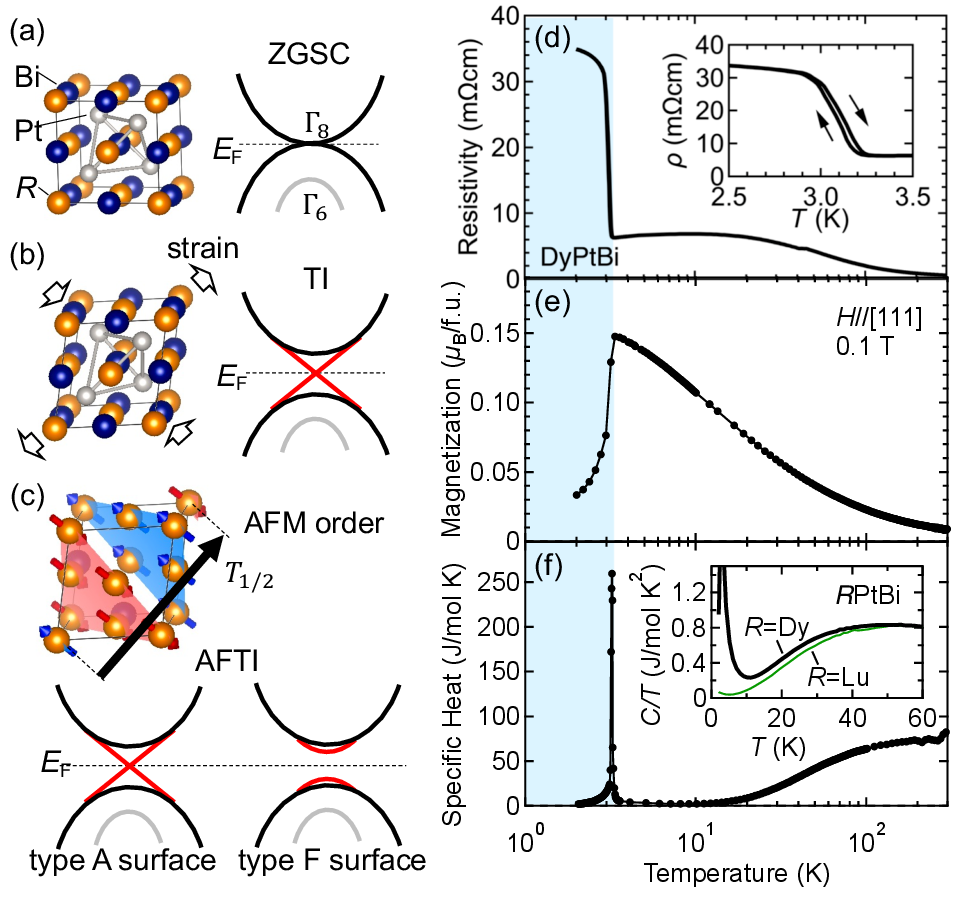}
\caption{\label{fig1}
Schematic picture of crystal structure and band structure for (a) zerogap semiconductor (ZGSC), (b) topological insulator (TI) induced by strain, and (c) antiferromagnetic topological insulator (AFTI) accompanied by the antiferromagnetic order.
The white arrows in (b) indicate the strain direction. The red and blue arrows in (c) denote the magnetic moments, and $T_{\rm 1/2}$ is a primitive-lattice translational vector.
Temperature ($T$) dependence of (d) resistivity, (e) magnetization, and (f) specific heat ($C$) for DyPtBi.
The inset of (d) shows the magnified view of resistivity around the transition temperature. The arrows indicate the resistivity on warming or cooling processes.
The inset of (f) shows the temperature dependence of $C/T$ for DyPtBi and nonmagnetic LuPtBi.
}
\end{figure}

The realization of AFTI was initially proposed in half-Heusler compounds which crystallize in a zinc-blende like structure~\cite{1991JAPCanfield,2016NPSuzuki,2016NMHirschberger,2018PNASShekhar,2023NCUeda,2025PRBUeda}.
Among them, paramagnetic $R$PtBi ($R$ being a rare-earth element) is a zero-gap semiconductor (ZGSC) with an inverted band structure analogous to HgTe~\cite{2006ScienceBernevig} [Fig.\ref{fig1}(a)].
Owing to the band-touching node protected by the cubic symmetry, ZGSC provides key ingredients for versatile topological states; for example, it transforms into a Weyl semimetal under magnetic fields~\cite{2016NPSuzuki,2016NMHirschberger,2018PNASShekhar} or into a topological insulator by strain [Fig.~\ref{fig1}(b)] as reported in mercury chalcogenides~\cite{2008PRBDai,2011PRLBrune,2013PRBWinterfeld}.
In many half-Heusler compounds, the rare-earth $4f$ magnetic moments order antiferromagnetically with the magnetic propagation vector $\boldsymbol{q}=(0.5,0.5,0.5)$ or equivalent directions at low temperatures.
This order breaks the time-reversal symmetry while preserves the $S=\Theta T_{\rm 1/2}$ symmetry, satisfying the conditions for AFTI [Fig.~\ref{fig1}(c)]~\cite{2010PRBMong,1991JAPCanfield}.
However, a feature of AFTI has not been reported so far in $R$PtBi.
This is presumably because the spin ordering alone yield a tiny or no bulk gap. For instance, the recent spectroscopic studies on GdPtBi reveal that it hosts tiny hole and electron Fermi surfaces near the center of Brillouin zone~\cite{2018PRLHutt,2023PRBPolatkan}.
Or otherwise, the presence of multiple magnetic domains may obscure intrinsic responses.

In this study, we explore such a hidden feature in the candidate material DyPtBi by means of the transport, ultrasonic, and neutron diffraction experiments combined with the uniaxial stress effects.
Dysprosium compounds in general have been proven to exhibit a variety of interesting magnetism~\cite{1999NatureRamirez,2004PRLGoto,2019PRBGao,2024NCAkatsuka}.
Among them, the electric quadrupoles of Dy $4f$ electrons, arising from the orbital degrees of freedom, show remarkable physical properties including possible electronic chirality~\cite{2026PRLIshitobi,2018PRBIshii,2025NCKurumaji} and structural phase transitions induced by ferroquadrupolar order~\cite{2005JPSJWatanuki,2007PRLJi,2018JPSJIshii} or cooperative Jahn-Teller effect~\cite{1972PRLMelcher,2010PRBKishimoto,1994JPSJNakamura,1997PBCMTakahashi}.
We find that, in DyPtBi, the transverse elastic constant $C_{44}$ shows the pronounced softening toward the antiferromagnetic transition temperature, indicative of the concomitant ferroquadrupolar order.
This order leads to the lattice distortion as well as the abrupt jump of resistivity, implying the opening of the charge gap followed by the degeneracy lifting of the band-touching node in ZGSC.
We reveal that the magnetic domains are aligned under compressive uniaxial stress on the basis of coupling between $4f$-moment order and lattice distortion, giving rise to the further remarkable increase of the resistivity by 14 \% as compared with the original state with naturally abundant domain-walls.
In turn, this observation indicates the emergence of anomalous metallic states at magnetic domain walls, which may be attributed to the conductive or gapless surface states analogous to those of magnetically-doped topological insulators or topological crystalline insulators.
These findings underline the high tunability and functionality of topological electronic systems entangled with multipolar degrees of freedom.


\section{Results}
\subsection{Phase transition in DyPtBi}
\begin{figure}
\includegraphics[width=1.0\columnwidth]{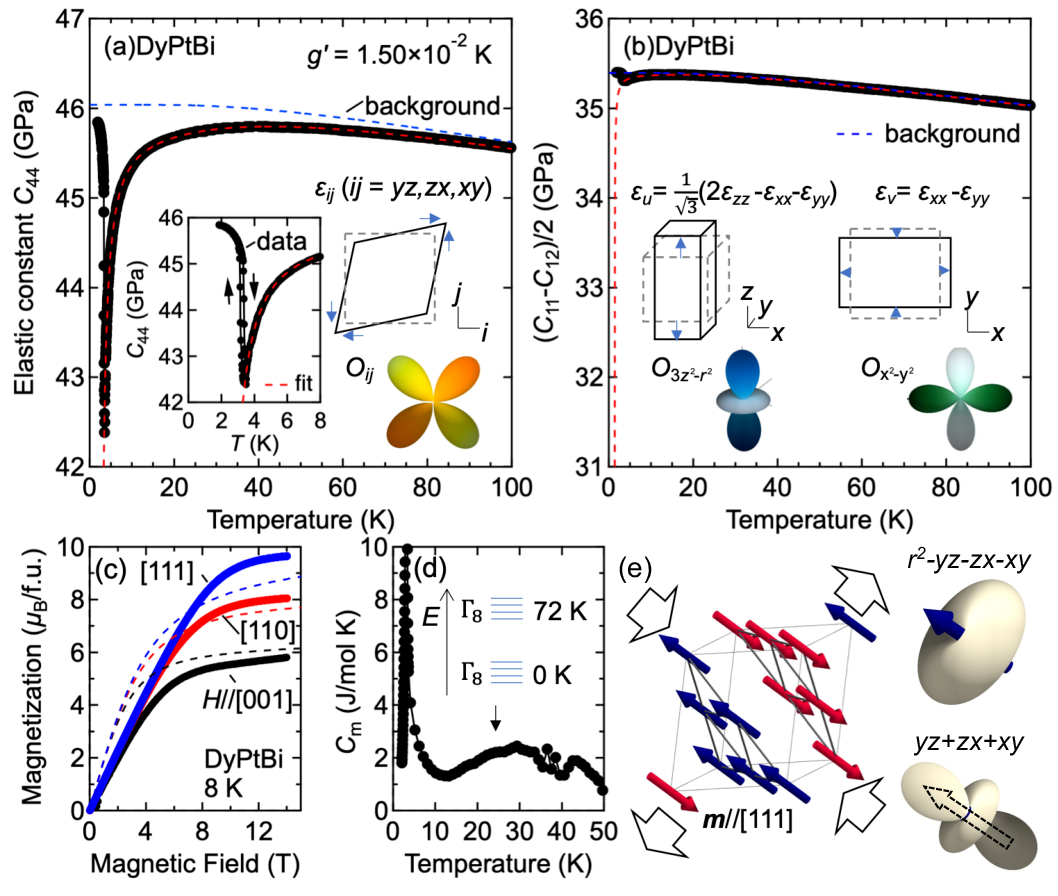}
\caption{\label{fig3}
Temperature dependence of elastic constants (a) $C_{44}$ and (b) $(C_{11}-C_{12})/2$ for DyPtBi. The insets show the magnified view of elastic constants, schematic pictures of the distorted lattices induced by the respective ultrasonic modes, and schematic pictures of multipole moments which couple to the respective modes. Dashed blue curves indicate the background and dashed red curves are theoretical fittings.
(c) Magnetic field dependence of magnetization for [111], [110], and [001] directions at 8 K. The dashed curves indicate the fitting curves.
(d) Temperature dependence of magnetic specific heat. The inset shows the crystal electric field level schemes for the ground state and first excited state.
(e) Schematic picture of antiferromagnetic and ferroquadrupolar order for DyPtBi.
The red and blue arrows represent the magnetic moments of the antiferromagnetic order, while the bold open (unfilled) arrows indicate the direction of the trigonal lattice distortion that emerges below the transition temperature.
}
\end{figure}

Figures~\ref{fig1}(d), (e), and (f) show the temperature ($T$) dependence of the resistivity, magnetization, and specific heat ($C$) in DyPtBi.
The resistivity exhibits a gradual increase upon cooling, reflecting the ZGSC with a low carrier density as illustrated in Fig.~\ref{fig1}(a).
Below 3.2 K, the resistivity shows an abrupt increase, accompanied by a thermal hysteresis between cooling and warming processes.
This resistivity jump in DyPtBi is distinct among other $R$PtBi compounds~\cite{1991JAPCanfield}, implying the formation of the significant charge gap, see Fig.~\ref{fig1}(b).
Nevertheless, the resistivity is apparently saturated at low temperatures, reminiscent of the Kondo insulator SmB$_6$ showing the surface electrical conductance, whose origin is under debate~\cite{2015PRLSyers,2020NRPLi}.
The magnetization exhibits Curie-Weiss-like temperature dependence at high temperatures and sudden decrease below the magnetic transition temperature $\tn =3.2$ K, as observed in other $R$PtBi compounds.
The previous study reveals that DyPtBi shows the antiferromagnetic order with the magnetic propagation vector $\boldsymbol{q}=(0.5,0.5,0.5)$~\cite{2020PRBZhang}.
Fitting to the Curie-Weiss law yields an effective magnetic moment of $10.9 \mu _{\mathrm{B}}$/mol and a Weiss temperature of $\theta _{\rm CW}=-15.2$ K.
The frustration index $f=|\theta _{\rm CW}|/\tn $ is 4.8, a typical value for general magnets~\cite{2022Mugiraneza} and rare-earth half-Heusler compounds~\cite{1991JAPCanfield,2025PRBUeda}.
The specific heat exhibits a sharp anomaly at $\tn $, characteristic of a first-order phase transition.
The inset of Fig.~\ref{fig1}(f) shows the temperature dependence of $C/T$ plotted for both DyPtBi and the nonmagnetic analog LuPtBi.
Since LuPtBi primarily reflects the phonon contribution, the deviation between DyPtBi and LuPtBi observed below 50~K is attributed to the magnetic contribution to the specific heat, and hereafter used as the magnetic specific heat $C_{\rm m}$ [Fig.~\ref{fig3}(d)] in DyPtBi.

\subsection{Ground state of Dy $4f$ electrons}
To gain an insight into the Dy $4f$ ground states, we performed ultrasonic measurements (for the measurement details, see Methods). This is a powerful technique for detecting electric multipolar moments which couple to local strain fields induced by ultrasound~\cite{text_Gschneidner}.
For instance, the elastic constant $C_{44}$ represents the linear response to the symmetric strains $\varepsilon_{yz}$, $\varepsilon_{zx}$, and $\varepsilon_{xy}$ with $\Gamma _{5}$ symmetry, which couple to the electric quadrupole moments $O_{yz}$, $O_{zx}$, and $O_{xy}$.
Similarly, $(C_{11}-C_{12})/2$ corresponds to the response to the symmetric strains $\varepsilon_u = (2\varepsilon_{zz} - \varepsilon_{xx} - \varepsilon_{yy})/\sqrt{3}$ and $\varepsilon_v = \varepsilon_{xx} - \varepsilon_{yy}$ with $\Gamma _{3}$ symmetry, which couple to the electric quadrupoles $O_{3z^2-r^2}$ and $O_{x^2-y^2}$.
These relationships are illustrated in the right insets of Figs.~\ref{fig3}(a) and \ref{fig3}(b).
In other words, the elastic constants can be regarded as the electric quadrupolar susceptibility with respect to the local strain fields, analogous to the magnetic susceptibility to the magnetic fields (see Methods in details).
The main panels of Figs.~\ref{fig3}(a) and \ref{fig3}(b) show the temperature dependence of $C_{44}$ and $(C_{11}-C_{12})/2$.
As the temperature decreases, $C_{44}$ increases monotonically, and then turns to decrease below 40 K.
A pronounced softening of 6 \% is observed towards $\tn$, characteristic of strong quadrupolar interactions.
With further cooling below $\tn $, $C_{44}$ shows abrupt hardening accompanied by thermal hysteresis, in accordance with the resistivity behavior observed in Fig.~\ref{fig1}(d).
On the other hand, $(C_{11}-C_{12})/2$ exhibits minimal softening towards $\tn $ [Fig.~\ref{fig3}(b)].
These results suggest that the electric quadrupoles with $\Gamma _{5}$ symmetry, $i.e.,$ $O_{yz}$, $O_{zx}$, and $O_{xy}$, play a central role in the phase transition at $\tn $.

We performed a theoretical fitting of the magnetization and magnetic specific heat $C_{\rm m}$ as well as the elastic constants for further analysis of crystalline electric field (CEF).
The details of the theoretical model are described in Methods.
Both elastic constants $C_{44}$ and $(C_{11}-C_{12})/2$ above $\tn $ are well reproduced by the fitting as depicted by the red dashed curves shown in Figs.~\ref{fig3}(a) and (b).
The deviation below $\tn $ reflects the critical softening of the elastic constant near the transition as the fitting is for non-ordered states.
In particular, the coupling constant of the quadrupolar intersite interaction, $g'_i$, is found to be positive, indicating that the ferroquadrupolar order is concomintantly accompanied by the antiferromagnetic transition, as often seen in Dy compounds~\cite{2018JPSJIshii}.
Figure \ref{fig3}(c) exhibits the magnetic field dependence of magnetization at 8 K ($>\tn $) along several field directions.
Each magnetization curve exhibits a monotonic increase following a Brillouin-like function but seemingly saturates at a different value, indicative of significant magnetic anisotropy.
The magnetization along the [111] direction reaches approximately 10 $\mu_{\rm B}$/mol at 14 T, which is close to the expected value for free ions.
On the other hand, the magnetization along the [001] direction is about half of that value.
The saturated experimental values obtained by extrapolation to the high-field limit are in good agreement with the calculated values (see Supplementary Note~I and Fig.~S1 in Supplemental Material).
The deviation of the fitting curves at the intermediate fields is likely due to the absence of interaction terms, such as antiferromagnetic exchange interactions, in the model.
Figure~\ref{fig3}(d) shows the temperature dependence of the magnetic specific-heat $C_{\rm m}$, which exhibits not only a sharp peak at $\tn $ but also a broad hump centered around 30 K.
This hump is attributed to the Schottky contribution between the ground-state $\Gamma_8$ quartet and the excited $\Gamma_8$ quartet lying above 72 K [inset of Fig. \ref{fig3}(d)].
In fact, the estimated magnetic entropy below 50 K is 16.3 J/K mol, which is reasonably close to the value of $R\ln8 \simeq 17.3$ J/K mol ($R$ being the ideal gas constant).

On the basis of these results, we discuss the ordered state of the Dy $4f$ electrons.
As discussed later, the compressive uniaxial stress applied along [111] ($\boldsymbol{\sigma }//[111]$) markedly enhances $\tn$, suggesting that the quadrupolar order is accompanied by a lattice distortion along the [111] direction. Thus, the electric quadrupoles, which linearly couple to the lattice distortion, can be associated with either an elongation or a compression of the $4f$ charge distribution along [111], as shown in the right panel of Fig.~\ref{fig3}(e).
The magnetic moment $\boldsymbol{m}$ has easy axes along $\langle 111\rangle $ directions of the parent cubic lattice, as observed in Fig.~\ref{fig3}(c).
As demonstrated later, we also find that $\boldsymbol{\sigma }//[111]$ suppresses the magnetic domain whose propagation vector $\boldsymbol{q}$ is $(111)$.
Assuming that the principal axis of the quadrupolar charge distribution is parallel to the magnetic easy axis owing to the strong coupling between quadrupolar and magnetic degrees of freedom, $\boldsymbol{\sigma }//[111]$ stabilizes the domain with $\boldsymbol{m}//(111)$.
Because $\boldsymbol{m}$ is not parallel to $\boldsymbol{q}$ in this system~\cite{2020PRBZhang}, $\boldsymbol{\sigma }//[111]$ effectively suppresses the $\boldsymbol{q}//(111)$ domain.
%
Figure~\ref{fig3}(e) illustrates the ordered state of the Dy $4f$ electrons.
Although the sign of the quadrupole-strain coupling coefficient $g$ cannot be determined (see Methods), the ultrasonic results constrain the possible ordered states to the two representative configurations illustrated in the right panel of Fig.~\ref{fig3}(e), both of which are described by linear combinations of the $O_{yz}$, $O_{zx}$, and $O_{xy}$ quadrupolar moments.
%
The ferroic order of quadrupolar moments results in a macroscopic lattice deformation that lowers the crystal symmetry, while the magnetic moments align antiferromagnetically along $\boldsymbol{q}$, with $\boldsymbol{q}\nparallel \boldsymbol{m}$.
We note that a previous neutron scattering study proposed an in-plane orientation of the magnetic moments in the (111) plane. Since the (111) plane does not contain any of the $\langle111\rangle$ axes, such an in-plane spin orientation appears inconsistent with the $\langle111\rangle$ easy-axis anisotropy. However, while the reported experimental data indicate that the magnetic moments are not perpendicular to the (111) plane, they do not necessarily establish that the moments lie strictly within the plane. The magnetic structure shown in Fig.~\ref{fig3}e is therefore consistent with the experimental data reported in Ref.~\cite{2020PRBZhang}.

\begin{figure}
\includegraphics[width=0.85\columnwidth]{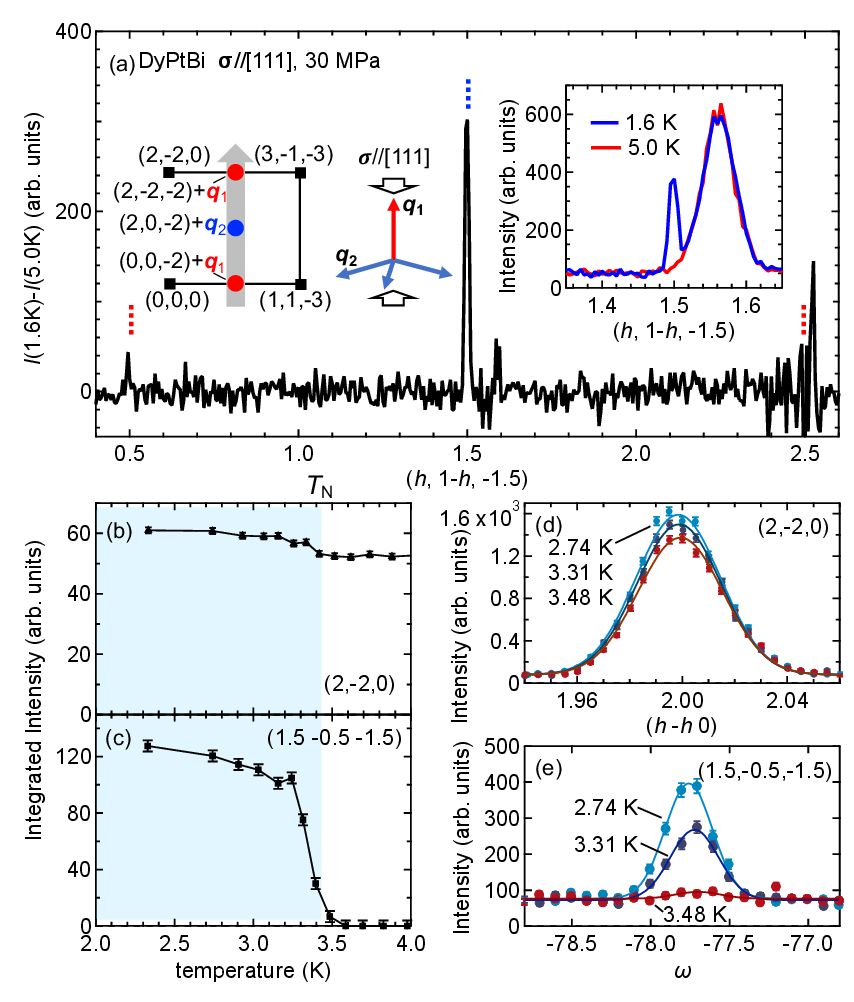}
\caption{\label{fig2}
(a) Difference of neutron scattering intensity between 1.6 K and 5.0 K scanning along the $(h,1-h,-1.5)$ reciprocal-lattice direction.
The left inset shows the scattering geometry where $\boldsymbol{q_1}$ and $\boldsymbol{q_2}$ denotes the magnetic propagation vectors. $\boldsymbol{q_1}$ is parallel to the uniaxial stress while $\boldsymbol{q_2}$ is not.
The right inset shows the neutron scattering at 1.6 K and 5.0 K.
Temperature dependence of neutron integrated intensity of (b) the nuclear peak $(2,-2,0)$ and (c) the magnetic peak $(1.5,-0.5,-1.5)$.
Blue-hatched regions indicate the antiferromagnetic order below $\tn $.
Neutron scattering of (d) $(2,-2,0)$ and (e) $(1.5,-0.5,-1.5)$ under uniaxial stress along the [111] direction at several temperatures.
Error bars indicate the standard error derived from the square roots of the numbers of counts. The curves are gaussian fit to experimental data.
}
\end{figure}

\subsection{Magnetic domain alignment by uniaxial stress}
Next, toward the zero-field control of the antiferromagnetic domains, we performed neutron scattering measurements under compressive uniaxial stress of 30~MPa along the [111] crystalline direction.
A previous study shows that the magnetic propagation vector $\boldsymbol{q}$ in DyPtBi is $\langle 0.5,0.5,0.5 \rangle $~\cite{2020PRBZhang}.
Owing to the cubic symmetry, there are four types of domains with different $\boldsymbol{q}$, which we refer to as q-domains.
Meanwhile, the easy axis of magnetic moments $\boldsymbol{m}$ is $\langle 111\rangle $, which we refer to as m-domains.
Since $\boldsymbol{m}$ along one of the $\langle 111\rangle$ axes on specific q-domain cannot be parallel to $\boldsymbol{q}$~\cite{2020PRBZhang}, there are twelve (four q-domains times three m-domains) types of magnetic domains in total.
The application of the uniaxial stress along the $[111]$ direction is thus expected to affect the volume fractions of the magnetic domains.

Here we selected the horizontal scattering plane on which magnetic Bragg peaks belonging to two of the q-domains, specifically $\boldsymbol{q_1}=(0.5,0.5,0.5)$ and $\boldsymbol{q_2}=(-0.5,-0.5,0.5)$, are accessible; the former is parallel to $\boldsymbol{\sigma }$ and the latter is not, as shown in the left inset of Fig.~\ref{fig2}(a).
The sample was essentially kept in its as-grown shape, with minimal polishing, in order to avoid introducing unintended strain and to maximize the neutron-scattering intensity.
Under these conditions, the domain populations are expected to be approximately balanced.
We note that $\tn $ is highly sensitive to uniaxial stress, as shown in the next section. Therefore, a small imbalance of the domain populations due to residual strain introduced during sample mounting cannot be completely excluded. Nevertheless, the previous observation of comparable intensities for all q-domains~\cite{2020PRBZhang} suggests that any such imbalance is sufficiently small for the purpose of examining the stress-induced evolution of the domain populations.
We performed reciprocal lattice scans along the $(h, 1-h, -1.5)$ line at 1.6 K and 5.0 K, below and above $\tn $, respectively.
Figure~\ref{fig2}(a) shows the difference in intensity between the two scans.
While a magnetic Bragg peak, indexed as $(2,0,-2)+\boldsymbol{q_2}$, is clearly observed at $(1.5,-0.5,-1.5)$, as indicated by a vertical blue dotted line, the magnetic reflections for the $\boldsymbol{q_1}$ domains are absent at the anticipated positions, indicated by vertical red dotted line.
The right inset of Fig.~\ref{fig2}(a) shows the raw data, where, despite the substantial background due to the stress cell, the magnetic peak of the $\boldsymbol{q_2}$-domain was clearly observed only below $\tn $.
In contrast, there are no temperature dependence at $h=0.5$ and $2.5$ which correspond to $\boldsymbol{q_1}$ (for details, see Supplementary Note II and Fig. S2).
These results unambiguously indicate that the application of the compressive uniaxial stress along $[111]$ completely suppresses the $\boldsymbol{q_1}$-domain.

Figures~\ref{fig2}(b) and (c) show the temperature dependence of the integrated intensities of the nuclear and magnetic Bragg reflections at $(2,-2, 0)$ and $(1.5,-0.5,-1.5)$ under the same uniaxial stress, respectively.
Figures~\ref{fig2}(d) and (e) show the rocking curve profiles for these reflections at selected temperatures.
These data were measured on heating, and the intensity of the magnetic reflection is reduced to zero above $\tn $.
The intensity of the nuclear reflection also shows a slight decrease at the same temperature, which could be ascribed to the lattice-structural transition at $\tn $ due to the afore-mentioned coupling between $4f$-moment order and lattice.
The antiferromagnetic order of this system could lead to a symmetry-lowering structural distortion below $\tn $.
Although the distortion itself could not be directly resolved by the present neutron scattering measurements, a slight change in crystal mosaicity can change the nuclear scattering intensity through the extinction effect.
These results are consistent with those reported in Ref.~\cite{2020PRBZhang}, suggesting that the application of the uniaxial stress does not largely affect the magnetic ground state of this system, but does the volume fractions of the respective magnetic domains.
It is noteworthy that $\tn $ slightly increases by 0.2 K with applying uniaxial stress of 30 MPa, which is also detected in magnetization and resistivity measurements as discussed later. 

There are several mechanisms for the domain alignment with applying uniaxial stress.
Firstly, as shown in Figs.~\ref{fig2}(b,c), $\tn $ markedly increases by approximately 6 \%, implying that the mechanism is not due to magnetocrystalline anisotropy~\cite{2025arxivZhaoyu} but magnetostriction~\cite{2012JPSJNakajima,1978Radhakrishna} or change of exchange interactions~\cite{2018JPSJNakajima,2011PRLRamazanoglu,2026PRXNeves} through the strong magnetoelastic coupling.
The compression along [111], in which ferromagnetic planes are stacked antiferromagnetically, may raise the magnetic energy and hence make the domain with $\boldsymbol{q_1}$ [see the middle inset of Fig.~\ref{fig2}(a)] energetically unfavorable.
More plausible scenario is related to the electric quadrupolar degrees of freedom as discussed in the previous section. Conventional magnetostriction originates from spin-lattice coupling and is generally proportional to the square of the magnetic order parameter. In contrast, quadrupolar moments represent anisotropic charge distributions and can couple directly to lattice strain through a bilinear interaction.
Thus, quadrupolar-lattice coupling can generate a significantly larger lattice response and domain-selective energy under relatively small applied stress. In fact, as demonstrated later, quite small uniaxial stress markedly increases $\tn$. Further, the lattice distortion has not been reported in other members of the $R$PtBi family, indicative of the importance of the quadrupolar-lattice coupling rather than spin-lattice coupling.

\begin{table}[b]
\caption{\label{tab:table1}%
Magnetic domains expected to survive under uniaxial stress, $\boldsymbol{\sigma }//[111]$ or $\boldsymbol{\sigma }//[1\overline{1}0]$. $\boldsymbol{q}$ is a magnetic propagation vector and $\boldsymbol{m}$ is a magnetic moment. There are twelve types of magnetic domains at ambient pressure except for the $\boldsymbol{m}//\boldsymbol{q}$ domains, which are energetically unfavored even without the uniaxial stress, denoted by hyphen. The domains denoted by ``$\boldsymbol{\sigma }//[111]$'' (``$\boldsymbol{\sigma }//[1\overline{1}0]$'') survive under uniaxial stress along [111] ([$1\overline{1}0$]) direction. The domains denoted by ``$\times $'' are eliminated by both uniaxial stress.
}
\begin{ruledtabular}
\begin{tabular}{lcccc}
\multicolumn{1}{c}{\textrm{}}&
\textrm{$\boldsymbol{m}$//$(111)$}&
\textrm{$\boldsymbol{m}$//$(\overline{1}\overline{1}1)$}&
\textrm{$\boldsymbol{m}$//$(\overline{1}1\overline{1})$}&
\textrm{$\boldsymbol{m}$//$(1\overline{1}\overline{1})$}\\
\colrule
$\boldsymbol{q}$//$(111)$ & - & $\times $ & $\boldsymbol{\sigma }//[1\overline{1}0]$ & $\boldsymbol{\sigma }//[1\overline{1}0]$ \\
$\boldsymbol{q}$//$(\overline{1}\overline{1}1)$ & $\boldsymbol{\sigma }//[111]$ & - & $\boldsymbol{\sigma }//[1\overline{1}0]$ & $\boldsymbol{\sigma }//[1\overline{1}0]$ \\
$\boldsymbol{q}$//$(\overline{1}1\overline{1})$ & $\boldsymbol{\sigma }//[111]$ & $\times $ & - & $\times $ \\
$\boldsymbol{q}$//$(1\overline{1}\overline{1})$ & $\boldsymbol{\sigma }//[111]$ & $\times $ & $\times $ & - \\
\end{tabular}
\end{ruledtabular}
\end{table}

\subsection{Anomalous metallic states at magnetic interfaces}
\begin{figure}
\includegraphics[width=0.95\columnwidth]{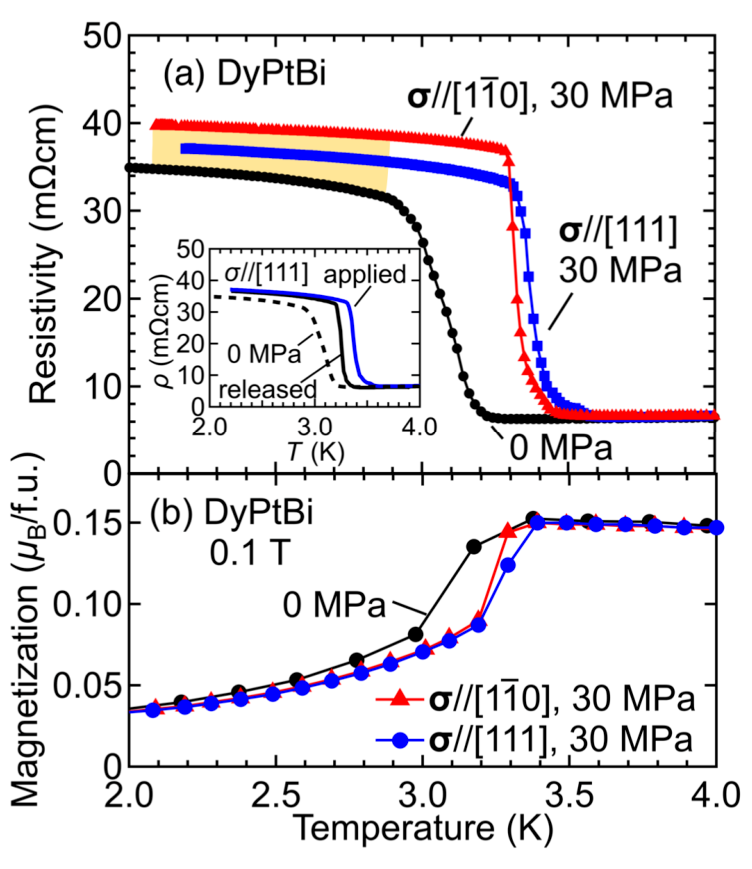}
\caption{\label{fig4}
Temperature dependence of (a) resistivity and (b) magnetization at ambient pressure and under uniaxial stress of 30~MPa along $[111]$ and $[1\Bar{1}0]$ directions.
Yellow-hatched region in (a) exemplifies the resistance increase (conductance decrease) due to the uniaxial strain-induced elimination of the gapless (conducting) domain wall states.
The inset in (a) shows the temperature dependence of resistivity at ambient pressure, under uniaxial stress of 30~MPa applied along $[111]$, and after the uniaxial stress was released. See the main text for details.}
\end{figure}

Finally, we investigate the uniaxial stress effect on the electrical transport and magnetization properties; the observed results are shown in Figs.~\ref{fig4}(a) and (b).
We applied compressive uniaxial stress of 30~MPa along both the [111] and [1$\Bar{1}$0] crystallographic directions using the same sample with the current along [1$\Bar{1}$0] direction.
One can see that $\tn $ slightly increases by $\sim 0.2$~K under uniaxial stress.
This enhancement is also discerned in the neutron scattering result [Fig.~\ref{fig2}(c)].
Notably, the resistivity below $\tn $ increases by approximately 7 \% for $\boldsymbol{\sigma }//[111]$ and 14 \% for $\boldsymbol{\sigma }//[1\Bar{1}0]$ compared to that at ambient pressure, despite no discernible change above $\tn $.
One might attribute it to the stabilization of lattice distortion under uniaxial stress, which could enlarge the bulk energy gap. However, such a scenario may lead to a larger increase of the resistivity under $\boldsymbol{\sigma }//[111]$ than that under $\boldsymbol{\sigma }//[1\Bar{1}0]$ because of the higher $\tn $; this is inconsistent with the observed result in Fig.~\ref{fig4}(a). Rather, the observed resistivity enhancement is more plausibly attributed to the change of the magnetic domain states.
As shown in the neutron diffraction results in Fig.~\ref{fig2}(a), uniaxial stress suppresses the q-domains with $\boldsymbol{q}//\boldsymbol{\sigma }$.
Furthermore, the resistivity below $\tn$ keeps unchanged after releasing stress as shown in the inset of Fig.~\ref{fig4}(a), inconsistent with the bulk origin scenario.
We note that the uniaxial stress is not completely released, but a residual stress of $25$ kPa because the ZrO$_2$ piston stays in contact with the sample (see Methods for details). Despite being three orders of magnitude smaller than the applied stress, this residual stress still produces a relatively large increase in $\tn $, highlighting the remarkable sensitivity of the long-range ordered state to uniaxial stress.

In general, the presence of magnetic DWs scatter charge carriers and enhance the resistivity~\cite{2000PRLKlein}.
However, the resistivity in Fig.~\ref{fig4}(a) shows the opposite behavior; some of DWs, that were present at ambient pressure, should be eliminated by suppressing q-domains under the uniaxial stress. This in turn suggests that the magnetic DWs in present DyPtBi may facilitate the charge transport or lead to the metallic conduction.
This scenario consistently explains the magnetic field dependence of resistivity as well (see Supplemental Material Note).
Notably, once after the magnetic field is applied, the resistivity measured again at zero field becomes larger as the respective volumes of magnetic domains are modified. Since the bulk electronic state should remain unchanged at zero field, magnetic DWs are responsible for change in electrical transport.

Given that the magnetic domain states in DyPtBi are similar to those in prototypical antiferromagnets such as MnO and NiO with $\boldsymbol{q}=\langle 0.5,0.5,0.5\rangle$~\cite{1960JAPRoth}, there are mainly two types of DWs; q-DWs [Fig.~\ref{fig0}(e)] and m-DWs [Fig.~\ref{fig0}(f)], respectively.
Note that the DWs are assumed to be atomically thin because of the strong anisotropy of Dy $4f$ magnetic moments.
In the case of q-DWs, spins remain antiferromagnetically aligned within DWs, preserving the $S=\Theta T_{\rm 1/2}$ symmetry.
For instance, between domains with $\boldsymbol{q}//(111)$ and $\boldsymbol{q}//(1\Bar{1}1)$, possible DW planes include (110) and (001).
In contrast, m-DWs form parallel to the ferromagnetic planes, e.g., the ($111$) plane, which breaks the $S$ symmetry.
We now consider the magnetic domain configurations under $\boldsymbol{\sigma }//[111]$ and $[1\Bar{1}0]$.
Firstly, as shown in Fig.~\ref{fig2}(a), q-domains with $\boldsymbol{q}$ nearly parallel to $\boldsymbol{\sigma }$ appear energetically unfavorable.
Furthermore, due to the lattice deformation coupled to the anisotropic orbitals around the magnetic moments, m-domains with $\boldsymbol{m}$ nearly perpendicular to $\boldsymbol{\sigma }$ should be suppressed.
Thus, for $\boldsymbol{\sigma }//[111]$, there remain magnetic domains with $\boldsymbol{m}//(111)$ and three propagation vectors 
$\boldsymbol{q}//(\Bar{1}\Bar{1}1)$, $\boldsymbol{q}//(\Bar{1}1\Bar{1})$, and $\boldsymbol{q}//(1\Bar{1}\Bar{1})$, as listed in Table~\ref{tab:table1}. This is consistent with the result shown in Fig.~\ref{fig2}(a).
In this case, only q-DWs are expected to remain [Fig.~\ref{fig0}(e)].
On the other hand, for $\boldsymbol{\sigma }//[1\Bar{1}0]$, the surviving domains likely include those with $\boldsymbol{q}//(111)$ and $\boldsymbol{q}//(\Bar{1}\Bar{1}1)$ which are perpendicular to $\boldsymbol{\sigma }//[1\Bar{1}0]$, as well as those with $\boldsymbol{m}//(\Bar{1}1\Bar{1})$ and $\boldsymbol{m}//(1\Bar{1}\Bar{1})$ which are not perpendicular.
Consequently, several types of DWs are expected to remain, different from the case of $\boldsymbol{\sigma }//[111]$. For example, for $\boldsymbol{m}//(\Bar{1}1\Bar{1})$, q-DWs between $\boldsymbol{q}//(111)$ and $\boldsymbol{q}//(\Bar{1}\Bar{1}1)$ survive whereas those between $\boldsymbol{q}//(111)$ and $\boldsymbol{q}//(1\Bar{1}\Bar{1})$ diminish (see Table~\ref{tab:table1}). Instead, m-DWs between $\boldsymbol{m}//(\Bar{1}1\Bar{1})$ and $\boldsymbol{m}//(1\Bar{1}\Bar{1})$, which are eliminated by $\boldsymbol{\sigma }//[111]$, remain.
It is noteworthy that the resistivity for $\boldsymbol{\sigma }//[1\Bar{1}0]$ is somehow larger than that for $\boldsymbol{\sigma }//[111]$, despite the larger number of remaining domains.
This fact suggests that different conduction channels may be realized depending on the type of DWs, unlike those in magnetic TIs which only host chiral edge currents.
Moreover, antiphase boundaries, which are expected to host chiral edge currents [Fig.~\ref{fig0}(d)], are likely to persist even under both types of uniaxial stress, possibly contributing to the saturation of the resistivity at low temperatures.

We crudely estimate the contribution of DWs to electrical conduction by using a simple parallel conductance model; $\sigma _{\rm total}=\sigma _{\rm bulk}+\sigma _{\rm DW}$ where $\sigma _{\rm total}$ is the total conductivity, $\sigma _{\rm bulk}$ is the bulk conductivity, and $\sigma _{\rm DW}$ is the DW conductivity.
For simplicity, we use the conductivity of the ambient pressure for $\sigma _{\rm total}$ and that under $\boldsymbol{\sigma }//[1\Bar{1}0]$ for $\sigma _{\rm bulk}$.
Under this assumption, we estimate $\sigma _{\rm DW}=3.6$ $\Omega ^{-1}$cm$^{-1}$ for the ambient-pressure state, from which the gapless DW states are partly eliminated by the uniaxial stress $\boldsymbol{\sigma }//[1\Bar{1}0]$.
Note that this value sets the lower limit for $\sigma _{\rm DW}$, yet is much larger than that of pyrochlore Nd$_2$Ir$_2$O$_7$ ($\sigma _{\rm DW}=0.77$ $\Omega ^{-1}$cm$^{-1}$)~\cite{2014PRBUeda,2015PRLUeda}.
Pyrochlore iridates are the first class of materials theoretically proposed to host a Weyl semimetal state~\cite{2011PRBWan}.
Especially, in Nd$_2$Ir$_2$O$_7$, such a state can be realized just below $\tn $ due to the breaking of time-reversal symmetry~\cite{2018NCUeda}.
In the lower-temperature insulating phase of Nd$_2$Ir$_2$O$_7$, anomalous metallic states are observed at magnetic domain boundaries via transport~\cite{2015PRLUeda}, terahertz spectroscopy~\cite{2014PRBUeda}, microelectrode device combined with microwave impedance microscopy~\cite{2015ScienceMa}.
In particular, the sheet conductance of DWs is estimated as $\sim 1$ mS, which is comparable to the surface conductance of SmB$_6$~\cite{2015PRLSyers} and that of the two-dimensional electron gas at LaAlO$_3$/SrTiO$_3$ interfaces~\cite{2013NCAnnadi}.
Considering that the DWs in the present DyPtBi are not completely eliminated even under uniaxial stress, they can host even higher conductance.
The realization of such conductive states at magnetic DWs, which are intrinsically robust against surface contamination and controllable by external fields, is highly promising for future device architecture.

\section{Discussion}
\subsection{Origin of metallic domain walls}
We discuss the possible origin of DW conduction.
One might consider that the bulk gap is locally suppressed at domain boundaries where the order parameter is disturbed, similar to Slater insulators or spin-density-wave states.
In such a case, the conductivity of the magnetic interface would be expected to be comparable to that of the paramagnetic bulk state above $\tn $.
Assuming a DW volume fraction of 1 \%, comparable to that reported for NiO~\cite{2018RMPBaltz}, the intrinsic conductivity of the DW region would be approximately $360$ $\rm \Omega ^{-1}cm^{-1}$. This value exceeds by more than a factor of two the bulk conductivity in the paramagnetic state above $\tn $, $166$ $\rm \Omega ^{-1}cm^{-1}$.
We note that this value is still likely underestimated.
First, the real $\sigma _{\rm DW}$ should be larger since the present uniaxial stress partially aligns the magnetic domain as shown in Table~\ref{tab:table1}. Besides, antiphase boundaries (Fig.~\ref{fig0}(d)) are not expected to be removed by either uniaxial stress or magnetic field.
Furthermore, as shown in Fig.~3(c), the magnetic anisotropy is substantial and hence the DWs can be much thinner than those in NiO, implying that the actual DW volume fraction is likely smaller than the assumed value. A smaller volume fraction would require an even larger DW conductivity.
Therefore, it is difficult to attribute the enhanced conductivity solely to the destruction of the ordered state while preserving the underlying electronic structure.

In magnetically doped TIs, chiral edge currents show up at boundaries between magnetic domains with opposite magnetization directions, or equivalently, with different Chern numbers [Fig.~\ref{fig0}(a)]~\cite{2017ScienceYasuda}.
Since AFTI can be regarded as an alternating stack of two-dimensional quantum Hall layers with opposite spin orientations as shown in Fig.~\ref{fig0}(c), the similar gapless states may be realized at antiphase boundaries [Fig.~\ref{fig0}(d)], and possibly at DWs in DyPtBi.
Another possibility is the symmetry breaking at the domain boundary.
For instance, SnTe, which is a TCI and characterized by mirror Chern numbers, undergoes a lattice distortion along the $[111]$ or equivalent directions at low temperatures.
This rhombohedral distortion eliminates a mirror plane perpendicular to the distorted direction, opening a mass gap in surface states.
Interestingly, each ferroelectric domain breaks crystal symmetry in a different direction.
Thus, depending on the ferroelectric domains, the Dirac masses at opposite momenta can change sign across a domain boundary, resulting in gapless states bound to DWs as shown in Fig.~\ref{fig0}(b)~\cite{2012NCHsieh}.
A similar situation can occur in AFTI, where $\boldsymbol{q}$ domains play a role analogous to ferroelectric domains in TCI [Fig.~\ref{fig0}(e)].
Moreover, AFTI offers an even richer platform than TCI, because the spin orientation in DWs provides an additional degrees of freedom.
This allows for a wider variety of theoretically proposed conduction channels, including spin-polarized flat bands~\cite{2021PRBPetrov} and topologically protected zero-line modes~\cite{2023PRBLiang}, which may be relevant to our observation in DyPtBi.

\subsection{Material design for antiferromagnetic topological insulators}
The antiferromagnetic ordering doubles the lattice periodicity along, for instance, the $\boldsymbol{a_3}$ direction which is one of the principal axes, leading to Brillouin-zone folding that the primitive $k_3=0$ and $k_3=\pi$ planes collapse onto the new $k_3^d=0$ plane, where $k_3$ is defined by $\boldsymbol{k}\cdot \boldsymbol{a_3}$.
As discussed in the original theoretical proposal of AFTIs~\cite{2010PRBMong}, the topological invariant of the folded band structure is determined by the sum of the invariants of the original planes. Therefore, when a strong TI with a nontrivial invariant at $k_3=0$ undergoes antiferromagnetic ordering, the new invariant $\gamma_0^d$ also remains nontrivial. As long as the combined symmetry of time-reversal and a half-lattice translation is preserved, as in antiferromagnets with q=(0.5,0.5,0.5), a $Z_2$ invariant can still be defined, and the system retains its topological character as an AFTI.

In the present system, as illustrated in Figs. 2(a) and 2(b), ZGSC can be transformed into a strong TI by opening a gap through symmetry breaking~\cite{2010NMChadov,2010NMLin}. In DyPtBi, the quadrupolar order breaks the crystal symmetry and opens such a gap, while the subsequent antiferromagnetic ordering realizes the AFTI state. Notably, the original AFTI proposal suggested that compounds such as GdPtBi, which do not exhibit a symmetry-lowering structural transition, might also host an AFTI phase. However, experimental evidence for such a scenario has not been obtained.
Our study highlights that, the cubic symmetry breaking, by employing multipolar orders that strongly couples to lattice degrees of freedom, or alternatively by inducing epitaxial strain in thin films, provides a promising route toward realizing AFTIs in band-inverted ZGSC half-Heusler compounds.

Finally, we discuss broader material-design strategies for AFTIs.
One promising approach is to introduce a staggered magnetization into a known strong TI~\cite{2010RMPHasan,2010PRBMong}.
While ferromagnetic TIs have been extensively studied~\cite{2019NRPTokura}, antiferromagnetic counterparts deserve renewed attention. In addition, materials possessing electronic structures similar to DyPtBi, including half-Heusler compounds such as $R$AuPb~\cite{2010PRBSawai} and pyrochlore iridates~\cite{2011PRBWan,2025RPPTokura}, constitute attractive candidate systems. The latter are particularly promising because strong electron correlations can generate sizable bulk gaps ($\sim 0.4$ eV~\cite{2016PRBUeda}), making them unique platforms for exploiting gapless states at surfaces and magnetic DWs.

\section{Conclusion}
In conclusion, we have demonstrated the presence of anomalous metallic states at magnetic interfaces in the antiferromagnetic topological insulator candidate DyPtBi by combining transport, neutron, and ultrasonic measurements.
The resistivity shows an abrupt, thermally-hysteretic jump at the magnetic transition temperature, being distinct from other $R$PtBi.
Concurrently, the transverse elastic constant $C_{44}$ exhibits the marked softening by 6 \% toward the transition temperature.
It implies that the ferroquadrupolar order induces a lattice distortion, which lifts the degeneracy of the quadratic band-touching node and consequently opens the charge gap in the bulk state, as manifested by the abrupt resistivity increase upon the ferroquadrupolar order.
Furthermore, we have found that the compressive uniaxial stress aligns the magnetic domains while eliminating the domain walls, leading to the further pronounced increase of 14 \% in resistivity.
This, in turn, points to the existence of gapless states at antiferromagnetic domain walls in the original state prior to the application of uniaxial stress.
The estimated conductivity of domain walls is much larger than that of Nd$_2$Ir$_2$O$_7$ in which the sheet conductance of domain walls is comparable to that of the two-dimensional electron gas at LaAlO$_3$/SrTiO$_3$ interfaces.
These findings reveal that the electric quadrupolar order is strongly coupled to the underlying topological electronic structure in DyPtBi, offering a promising pathway toward deeper understanding and further development of composite topological materials.

\section{Methods}

DyPtBi single crystals were synthesized by the Bi flux growth method.
A nominal $R$PtBi$_{9}$ molar ratio of R (99.9 \%), Pt (99.99 \%), and Bi (99.999 \%) was loaded in an alumina crucible.
The crucibles were sealed under vacuum inside a silica ampule and heated to 1150 $^{\circ }$C in 12 h. After dwelling at 1150 $^{\circ }$C for 12 h, the furnace was slowly cooled in 100 h to 800 $^{\circ }$C, and then the ampule was rapidly inverted into a metal centrifuge and the excess flux decanted.
The crystal structure of single crystals was characterized by Rigaku x-ray diffraction system and the lattice constant was determined as $a=6.64$ \AA , consistent with the previous study~\cite{2020PRBZhang,2021AFMChen}.

The crystals were polished into rectangular samples with (111) surface for transport measurements using the standard four-probe method.
Gold wires are attached to the samples with silver epoxy.
The current direction is parallel to $[1\Bar{1}0]$ crystalline direction.
Resistivity, specific heat, and magnetization were measured using a Quantum Design physical property measurement system (PPMS) and magnetic property measurement system (MPMS).
To apply uniaxial stresss along [111] and $[1\Bar{1}0]$ directions during the transport and magnetization measurements, we employed a home-made uniaxial stress cell developed in Ref~\cite{2011PRBNakajima,2015NCNii}.

The elastic constants $C_{44}$ and $(C_{11}-C_{12})/2$ were measured between 2 and 150 K using a phase comparison-type pulse-echo method~\cite{1969PRMoran}.
The frequency of the ultrasound is $\sim 20$ MHz.
The propagation $\boldsymbol{k}$ and polarization $\boldsymbol{u}$ directions of the ultrasound are $\boldsymbol{k}//[001]$ and $\boldsymbol{u}//[110]$ for $C_{44}$, and $\boldsymbol{k}//[110]$ and $\boldsymbol{u}//[1\Bar{1}0]$ for $(C_{11}-C_{12})/2$, respectively.
The elastic constant $C$ was calculated using the relation $C=\rho v^2$, where $\rho =12.83$ g/cm$^3$ is the room-temperature mass density and $v$ is the sound velocity determined at 2 K using the sample length and a time interval between pulse echoes.

For crystalline electric field (CEF) analysis, we considered the following effective Hamiltonian assuming non-ordered state and the low frequency limit~\cite{1962Lea,1964Hutchings}:
$$\mathcal{H}_{\rm eff}=\mathcal{H}_{\rm CEF}+\mathcal{H}_{\rm strain}+\mathcal{H}_{\rm Q-Q}$$
$$\mathcal{H}_{\rm CEF}=W\left\lbrack x\left(\frac{O^{0}_{4}+5O^{4}_{4}}{F(4)}\right)+(1-|x|)\left(\frac{O^{0}_{6}-21O^{4}_{6}}{F(6)}\right) \right\rbrack$$
$$\mathcal{H}_{\rm strain}=-\sum_{\alpha }\sum_{\Gamma _{\gamma }}g_{\Gamma _{\gamma }}O^{\alpha }_{\Gamma _{\gamma }}\varepsilon_{\Gamma _{\gamma }}$$
$$\mathcal{H}_{\rm Q-Q}=\sum_{\alpha }\sum_{\Gamma _{\gamma }}g'_{\Gamma _{\gamma }}\langle O_{\Gamma _{\gamma }}\rangle O^{\alpha }_{\Gamma _{\gamma }}$$
where $W$ is the scale factor, $x$ is the ratio, $F(4)=60$ and $F(6)=13860$ are factors common to all the matrix elements for $J=15/2$, and $O^{n}_{m}$ are the Stevens equivalent operators~\cite{1962Lea}.
$g_{\Gamma _{\gamma }}$ is the strain-quadrupole coupling constant, $g'_{\Gamma _{\gamma }}$ is the coupling constant of the quadrupolar intersite interaction, and $O_{\Gamma _{\gamma }}$ is the quadrupole operator, which belong to the irreducible representation $\Gamma _{\gamma }$.
$H_{\rm strain}$ and $H_{\rm Q-Q}$ affect 4$f$-electronic states as a perturbation. 
The strain susceptibility is calculated by using the Wigner-Brillouin perturbation method. 
For the strain-quadrupole interaction, $H_{\rm strain}$, 
when the strain $\varepsilon_\Gamma$ is induced by ultrasound, 
each energy of the CEF state up to the second-order perturbation, $E_k$, is considered as 
\begin{eqnarray}
\displaystyle E_k(\varepsilon_\Gamma) = E^0_k 
- g_\Gamma \langle k| O_\Gamma |k \rangle \varepsilon_\Gamma 
+ g_\Gamma^2 \sum_{k \neq l} \frac{\left| \langle k| O_\Gamma |l \rangle 
\right|^2}{E^0_k-E^0_l} \varepsilon_\Gamma^2, \nonumber 
\end{eqnarray}
where $E^0_k$ is an energy of non-perturbed CEF state. 
The Helmholtz free energy contributed from 4$f$-electrons, $F_{\rm ion}$, 
is represented by following equation: 
\begin{eqnarray}
\displaystyle F_{\rm ion}(\varepsilon_\Gamma, T) &=& -N_0 k_{\rm B} T 
\ln Z (\varepsilon_\Gamma, T) \nonumber \\
&=& -N_0 k_{\rm B} T \ln \left[ \sum_k \exp \left( - 
\frac{E_k(\varepsilon_\Gamma)}{k_{\rm B}T} \right) \right], \nonumber 
\end{eqnarray}
where $Z$ is the partition function and $k_{\rm B}$ is the Boltzmann constant. 
The total free energy $F_{\rm total}$ is given by following equation: 
\begin{eqnarray}
F_{\rm total} = F_{\rm ion}+\frac{1}{2} C_0 \varepsilon_\Gamma^2, \nonumber 
\end{eqnarray}
where the second term is the free energy of phonons and electrons other than 4$f$-electrons. 
The elastic constant $C_{\rm SQ}$ for the strain-quadrupole interaction 
is defined by the second-order derivative of $F_{\rm total}$ with respect to $\varepsilon_\Gamma$: 
\begin{eqnarray}
\displaystyle C_{\rm SQ}(T) = \left( \frac{\partial^2 F_{\rm total}
(\varepsilon_\Gamma, T)}{\partial \varepsilon_\Gamma^2} \right)_{\varepsilon_\Gamma 
\rightarrow 0} = C_0 - N_0 g_\Gamma^2 \chi_s (T), \nonumber 
\end{eqnarray}
where $\chi_s$ is the strain susceptibility. 
$\chi_s$ is given by following equation: 
\begin{eqnarray}
\displaystyle -g_\Gamma^2 \chi_s (T) = \sum \left[ \left\langle \frac{\partial^2 E_k}{\partial \varepsilon_\Gamma^2} 
\right\rangle - \frac{1}{k_{\rm B} T} \left\{ \left\langle \left( 
\frac{\partial E_k}{\partial \varepsilon_\Gamma} \right)^2 \right\rangle - 
\left\langle \frac{\partial E_k}{\partial \varepsilon_\Gamma} \right\rangle^2 \right\} 
\right], \nonumber 
\end{eqnarray}
where $\langle ~ \rangle$ denotes the thermal average. 
In addition to the strain-quadrupole interaction, 
the quadrupole-quadrupole interaction, $H_{\rm Q-Q}$, is also added by using the mean-field approximation of $O_\Gamma$. 
Consequently, the temperature dependence of elastic constant $C_{ii}(T)$ 
including both $H_{\rm strain}$ and $H_{\rm Q-Q}$ is represented by the following equation: 
\begin{eqnarray}
\displaystyle C_{ii}(T) = \frac{-N_0 g_\Gamma^2 \chi_s (T)}
{1-g'_\Gamma \chi_s (T)} + C_0(T). \nonumber
\end{eqnarray}
where $N_{0}=1.364\times 10^{28}$ m$^{-3}$ is the number density of Dy ions per unit volume and $\chi _{s}$ is the strain susceptibility.
We adopted the Varshni equation for the background stiffness~\cite{1970PRBVarshni}:
$$C_{0}=C_{\rm 0K}-\frac{s}{\exp(t/T)-1},$$
where $C_{\rm 0K}$ is the elastic constant at 0 K, $s$ and $t$ are fitting parameters.
We use $x=-0.36$ and $W=-1.7$ K for the fitting.
Other fitting parameters for both $C_{44}$ and $(C_{11}-C_{12})/2$ are summarized in table~\ref{tab:ultrasonic}.
Both elastic constants $C_{44}$ and $(C_{11}-C_{12})/2$ are well reproduced by the fitting as depicted by the red dashed curves shown in Figs.~\ref{fig3}(a) and (b).

\begin{table*}[t]
\caption{\label{tab:ultrasonic}%
Fitting parameters for ultrasonic measurements. $g$ is the strain-quadrupole coupling constant, $g'$ is the coupling constant of the quadrupolar intersite interaction, $C_{\rm 0K}$ is the elastic constant at 0 K, $s$ and $t$ are fitting parameters.
}
\begin{ruledtabular}
\begin{tabular}{lccccc}
\multicolumn{1}{c}{\textrm{}}&
\textrm{$|g|$ (K)}&
\textrm{$g'$ (K)}&
\textrm{$C_{\rm 0K}$ (GPa)}&
\textrm{$s$ (GPa)}&
\textrm{$t$ (K)}\\
\colrule
$C_{44}$ & $13.3$ & $+2.15\times 10^{-2}$ & $46.1$ & $1.05$ & $125$ \\
$(C_{11}-C_{12})/2$ & $8.42$ & $+0.25$ & $35.4$ & $0.49$ & $85$ \\
\end{tabular}
\end{ruledtabular}
\end{table*}

Neutron scattering experiments were carried out on the polarized neutron triple-axis spectrometer (PONTA) at 5G beamline of Japan Research Reactor 3 (JRR-3) in Japan.
The single crystal sample with the dimensions of $1.4\times 1.4 \times 0.7$ mm$^3$ was mounted in a clamp-type uniaxial-stress cell to apply a compressive stress of 30~MPa along the $[111]$ direction, which is normal to the widest surfaces of the sample.
The cell was loaded into a pumped $^4$He cryostat using a tilting mount for the uniaxial stress cell, so that the direction of the uniaxial stress was tilted by $\sim 10^{\circ }$ from the vertical direction.
This setup enabled us to access magnetic reflections belonging to the $\boldsymbol{q_1}$ and $\boldsymbol{q_2}$ domains, as detailed in the main text.
The spectrometer was operated in the unpolarized two-axis mode with the horizontal beam collimation of open-80’-80’.
An incident neutron beam with the energy of 34.05 meV was obtained by a pyrolytic graphite (002) monochromator.

Neutron scattering measurements under magnetic fields in Supplemental Material Figs.~S1(c) and (d) were performed using the High-Resolution Chopper spectrometer (HRC) at BL12~\cite{2011Itoh} in the Materials and Life Science Facility of the Japan Proton Accelerator Research Complex (J-PARC).
A single crystal of DyPtBi with the dimensions of $2\times 2 \times 2$ mm$^3$ was loaded into a vertical-field superconducting magnet.
The direction of the magnetic field was parallel to the [111] direction of the sample.
We measured time-of-flight (TOF) neutron Laue diffraction patterns with a polychromatic pulsed neutron beam with varying magnetic field at the base temperature of 1.6 K.
Integrated intensities of selected reflections were extracted from the intensity maps on the two-dimensional position sensitive detectors.

\section*{Acknowledgments}

We are grateful to Masataka Mogi, Shunsuke Kitou, and Taka-hisa Arima for fruitful discussions.
The neutron scattering experiments at PONTA in JRR-3 were carried out along the proposals (No.~ 22517). The neutron scattering experiments at HRC in the MLF of J-PARC was performed under a user program (Proposal No. 2021S01).
This work was supported by JSPS Grant-in-Aid for Scientific Research (No.~23H05431, 24H01649, 25K00957, 25K22014), by The Fujikura Foundation, by the Inamori Foundation, and by the Murata Science and Education Foundation, Japan.

\bibliographystyle{apsrev4-2}
\bibliography{reference}

\end{document}